\documentclass[ reprint,showpacs,
 superscriptaddress,
 pra,
]{revtex4-1}
\usepackage{mathrsfs,amssymb}
\usepackage{amsmath}
\usepackage{bm}
\usepackage{graphics}
\usepackage{grffile}
\usepackage{empheq}
\usepackage{natbib}
\usepackage{cases}
\usepackage{mhchem}

\renewcommand{\vec}[1]{\bm{\mathrm{#1}}}
\begin{document}

\title{All optical quantum storage based on spatial chirp of the control field}
\author{Xiwen Zhang}
\email{xiwen@physics.tamu.edu}
\affiliation{Department of Physics and Astronomy and Institute for Quantum Studies, Texas
A\&M University, College Station, Texas 77843-4242, USA}
\author{Alexey Kalachev}
\affiliation{Zavoisky Physical-Technical Institute of the Russian Academy of Sciences,
Sibirsky Trakt 10/7, Kazan, 420029, Russia}
\affiliation{Kazan Federal University, Kremlevskaya 18, Kazan, 420008, Russia}
\author{Olga Kocharovskaya}
\affiliation{Department of Physics and Astronomy and Institute for Quantum Studies, Texas
A\&M University, College Station, Texas 77843-4242, USA}
\date{\today }

\begin{abstract}
We suggest an all-optical quantum memory scheme which is based on the off-resonant Raman interaction of a signal quantum field and a strong control field in a three-level atomic medium in the case, when the control field has a spatially varying frequency across the beam, called a spatial chirp. We show that the effect of such a spatial chirp is analogous to the effect of a controllable reversible inhomogeneous broadening (CRIB) of the atomic transition used in the gradient echo memory (GEM) scheme. However, the proposed scheme does not require temporal modulation of the control field or the atomic levels, and can be realized without additional electric or magnetic fields. It means that materials demonstrating neither linear Stark nor Zeeman effects can be used and/or materials which are placed in specific external fields remain undisturbed.
\end{abstract}

\maketitle


\section{Introduction}

In recent years a field of an optical quantum memory \cite{Hammerer10,Tittel10,Simon10} has undergone a rapid development as one of
the leading research fields in the whole quantum information endeavor. In particular, storage and retrieval of single-photon states, which are used as flying qubits, is of crucial importance for developing long-distance quantum communication via quantum repeaters and scalable optical quantum computing. The storage of the temporal profile of a single-photon wave packet has been
demonstrated using different techniques based, for example, on an electromagnetically induced transparency (EIT)~\citep{Chaneliere05, Eisaman05,Novikova07, Choi08, Heinze13, Chen13} and an off-resonant Raman interaction~\cite{Reim10, Reim11, Reim12}. In the EIT scheme, which requires less power than Raman, a dynamical control of the group velocity of the input signal pulse allows for a mapping of the signal field into an atomic coherence~\citep{Fleischhauer00}. While in the Raman scheme, a specially shaped-in-time control field provides a mode matching~\citep{Nunn07} which similarly maps the signal into the atomic Raman coherence. The Raman scheme generally allows for a broadband and low-noise quantum storage and is insensitive to an inhomogeneous broadening. However, in either one a dynamically shaped control field~\citep{Gorshkov07(0), Novikova08} and a synchronization of the control and signal fields are required. So, a prior knowledge on the signal pulse timing is needed in order to have it stored, which can strongly limit possible practical application of such schemes in the quantum information processing. The schemes based on the photon-echo effect, such as controlled reversible inhomogeneous broadening (CRIB)~\citep{Moiseev01, Nilsson05, Kraus06}, (in particular, gradient echo memory (GEM)~\citep{Hetet08}) and an atomic frequency comb (AFC)~\citep{Afzelius09} do not require modulation or synchronization of the control field with the input signal or can be implemented in a two-level system without any control field. In doing so, high efficiency~\cite{Hosseini11, Hedges10, Sabooni13}, bandwidth~\citep{Saglamyurek11}, and mode capacity~\citep{Usmani10,Bonarota11} of the protocols have been demonstrated. However, the GEM scheme can only be implemented in a material which demonstrates the Stark or Zeeman effect used to create a spatially dependent inhomogeneous broadening. In the AFC scheme a tailored inhomogeneous broadening (frequency comb) is employed to assist the storage of the incident signal, and such a periodic absorption spectrum generally needs a delicate preparation processing.

Recently, the quantum memory schemes based on phase matching control (PMC) were suggested~\citep{Kalachev11, Zhang13, Kalachev13, Zhang14LP, Clark12}. They demonstrate a longitudinal CRIB scheme performance without controlling inhomogeneous broadening. PMC quantum memory schemes typically use the Raman configuration, but unlike EIT or Raman scheme they do not require synchronization of the control and signal fields, and can operate with control field of a constant amplitude. However they imply a manipulation with time of the refractive index or control field propagation direction. In this paper we suggest a quantum storage scheme which also employs the Raman configuration but requires neither the synchronization of the fields, nor the use of inhomogeneous broadening, nor the time variation of the control field amplitude or any other parameters of the system. In this quantum memory scheme, the off-resonant Raman interaction is assisted by a spatial chirp of the control field. The spatial chirp implies a varying frequency of the control field across the beam. While it shares some similarities with the PMC quantum memory schemes, the proposed one does not involve any temporal manipulation of the control field and, thus, is more robust to a time jitter.

It is worth comparing the proposed scheme with the PMC quantum memory scheme via a temporal frequency chirp of the control field~\citep{Zhang14LP}. In the latter one the chirp generates time-dependent control-field wave vector which sweeps the phase matching condition and maps the temporal information of the signal filed into the spin grating. However, it produces a time-dependent detuning which drives the signal out of the two-photon Raman resonance during the process of storage. In order to compensate this detuning, an additional synchronous manipulation of the transition frequency of the medium is required, making the scheme difficult for implementation. While in the proposed scheme, the spatial chirp generates space-dependent control-field frequency to achieve the same effect but leaves the two-photon detuning stationary. This makes our scheme much simpler and robust.

The paper is organized as follows. In Sec. \ref{sec: The model and basic equations} we introduce a model which describes the proposed quantum memory scheme. In Sec. \ref{sec: Transverse excitation} we consider the case of a transverse excitation, when the propagation directions of the signal and control fields are perpendicular to each other, and show that it is analogous to the GEM scheme. In Sec. \ref{sec: Non-Transverse excitation} we discuss peculiarities of non-transverse excitation geometry. In Sec. \ref
{sec: Spatial chirp of Gaussian control beams} we discuss the experiment related spatial chirp of the control beams. Sec. \ref{sec:Discussion and Conclusion} elaborates on the essence of the proposed spatial-chirp memory scheme and concludes the paper.

\section{The model and basic equations}

\label{sec: The model and basic equations}

\begin{figure}[th]
\centering
\resizebox{!}{8.9cm}{\includegraphics{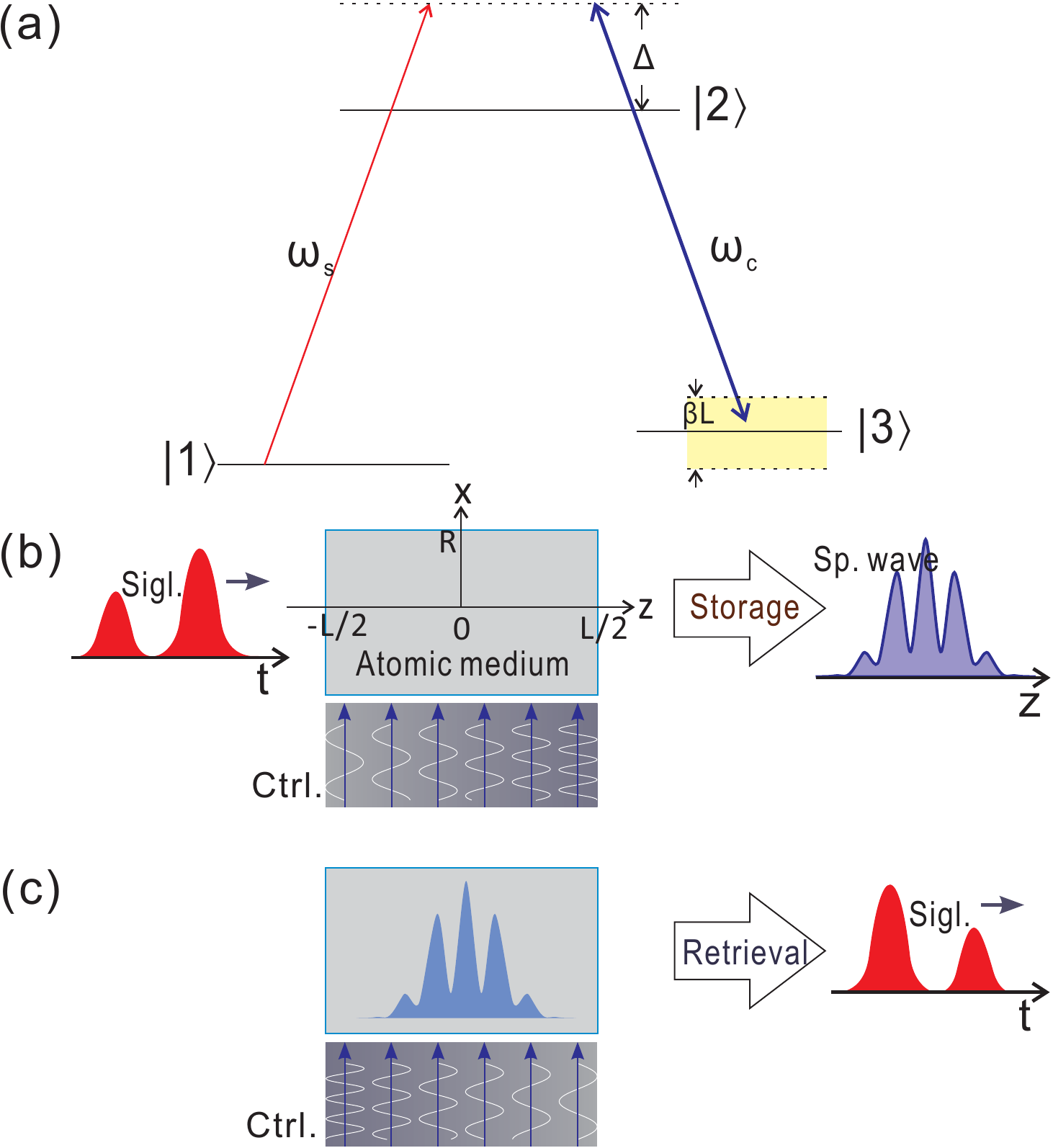}}
\caption{(Color online) (a) Energy diagram of the off-resonant Raman interaction in a three-level $\Lambda $ system. The angular frequencies of the signal and control fields are $\protect\omega _{s}$ and $\protect\omega_{c}$, respectively; $\Delta $ is the one-photon detuning between the fields and the corresponding atomic transitions. The control field spectrum width is given by the spatial chirp $\Delta \protect\omega _{c}$, which is $\protect\beta L$ for the transverse excitation setup. (b, c) Transverse
excitation setup in which the propagation directions of the signal and control fields are perpendicular to each other: The signal, propagating along the $z$ axis of the cylindrical medium, experiences only the longitudinal inhomogeneity of the Raman interaction with the control field that is spatially chirped along the $z$ axis. During storage (b), the temporal profile of the signal field is mapped into a pattern of the spin wave (Raman coherence) distribution due to the Raman interaction of the signal with the longitudinally chirped control field. During retrieval (c),the stored spin-wave generates the output signal field and the spin-wave spatial pattern is mapped back into the output-signal temporal profile. }
\label{Figure: 1Setup}
\end{figure}

We consider the off-resonant Raman interaction between a signal field (single-photon wave packet) $E_{s}$ and a three-level atomic ensemble under the control of a strong classical field $E_{c}$, see Fig. \ref{Figure: 1Setup}(a). The atoms, which have a $\Lambda$-type energy level structure, are assumed to be stationary and uniformly distributed in space over the sample volume. Let us put the coordinate origin at the center of the sample and define the longitudinal direction of the medium as $\hat{z}$. The excitation volume of the atomic medium is a cylinder with a length $L$ and radius $R$. We assume the intensities of the signal and control fields to be constant across the transverse to their propagation direction dimensions, and their polarizations to be orthogonal so that they are coupled to the different atomic transitions. The control field frequency $\omega _{c}$ has a linear
transverse spatial chirp. In the transverse excitation regime (Fig. \ref{Figure: 1Setup}(b, c)), the signal field propagates along $\hat{z}$ and the control field propagates perpendicularly to it. The storage of the single-photon wave packet takes place during an interval of time $t\in(-T,0) $, and the retrieval happens during $t\in (0,T)$.
The spatial chirp of the control field creates an effective inhomogeneous broadening $\beta L$ (Fig. \ref{Figure: 1Setup}(a)) on which the spectrum of the signal field is converted into the spatial spectrum of the spin wave (Raman coherence) during the storage process and back during the retrieval process. The storage time is defined by the Raman coherence decay time. But for the sake of simplicity, a free decay of the spin wave between the end of the storage and the beginning of the retrieval will be omitted.

The signal field (single-photon wave packet) of an average frequency $\omega_{s}$ and wave vector $k_{s}$ can be represented as
\begin{equation}
E_{s}(\vec{r},t)=\frac{i}{n}\sqrt{\frac{\hbar \omega _{s}}{2\varepsilon _{0}c}}a(\vec{r},t)\,e^{i(\vec{k}_{s}\cdot \vec{r}-\omega _{s}t)}+\text{H.c.,}
\label{Signal Field Def}
\end{equation}
via the slowly varying annihilation operator $a(\vec{r},t)$, where $c$ is the phase velocity of the signal field inside the medium, and $n$ is the refractive index which takes into account the contributions from the host material and the resonant atoms.

Assuming the classical control field has a large Fresnel number for the interaction volume, we write it as a plain wave with a spatial chirp across the beam:
\begin{equation}
E_{c}(\vec{r},t)=E_{0}e^{i[\vec{k}_{c}(\vec{r}_{\perp })\cdot \vec{r}-\omega_{c}(\vec{r}_{\perp })t-\varphi _{0c}(\vec{r}_{\perp })]}+\text{c.c.}.
\end{equation}
We will consider a linear transverse spatial chirp $\omega _{c}(\vec{r}_{\perp })=\omega _{c0}+\vec{\beta}\cdot \vec{r}_{\perp }$, where the radius-vector $\vec{r}_{\perp }$ is perpendicular to the propagation direction of the control field, $\omega_{c0}=\omega _{c}(\vec{r}_{\perp }=0)$, $\vec{\beta}$ is a frequency gradient. A transverse pattern of the wave vector of the control field is described by a vector function $\vec{k}_{c}(\vec{r}_{\perp }) \perp \vec{r}_{\perp }$, $|\vec{k} _{c}(\vec{r}_{\perp})|=\omega _{c}(\vec{r}_{\perp })/c$, and a phase shift $\varphi _{0c}(\vec{r}_{\perp })=\vec{k}_{c}(\vec{r} _{\perp })\cdot \vec{r}_{0}-\omega _{c}(\vec{r}_{\perp })t_{0}$ specifies the position $\vec{r}_{0}$ and time $t_{0}$ of the constant phase plane. We assume the control field is kept constant during the writing and retrieval processes and switched off and on instantaneously accordingly by the end of the writing and beginning of reading processes. In principle, switching rate faster than the inverse one-photon detuning could induce non-adiabatic excitation of optical coherence which has fast decay rate and opens a loss channel. However for sufficient large one-photon detuning compared to the Rabi frequency (in the example provided in Sec. \ref{sec: Spatial chirp of Gaussian control beams}, this ratio is $\approx 10$), such negative influence is negligible~\cite{Moiseev13}.

The collective atomic operators are defined as the density of the single-atom operators
\begin{equation}
\sigma _{mn}(\vec{r},t)=\frac{1}{N}\sum_{j}\lvert m_{j}\rangle \langle n_{j}\rvert \,\delta ^{(3)}(\vec{r}-\vec{r}_{j})\,,
\end{equation}
where $N$ is the constant atomic number density, and $\lvert n_{j}\rangle $ is the $n$th state ($n=1,2,3$) of $j$th atom with the energy $\hbar \omega_{n}$; the lowest energy level is $\omega _{1}=0$. Let us introduce the slowly varying collective spin (Raman coherence) operator $s\left( \vec{r},t\right) $ such that
\begin{equation}
\sigma _{13}\left( \vec{r},t\right) =s\left( \vec{r},t\right) \,e^{i\left(
\vec{k}_{s}-\vec{k}_{c0}\right) \cdot \vec{r}-i(\omega _{s}-\omega
_{c0})t}\,.  \label{Def s}
\end{equation}
The off-resonant Raman interaction can be described by the following equations:
\begin{align}
& \bigg(\frac{\partial }{\partial z}+\frac{1}{c}\frac{\partial }{\partial t}%
\bigg)a(\vec{r},t)=i\frac{\Delta _{\perp }}{2k_{s}}a(\vec{r},t)-g^{\ast }Ns(%
\vec{r},t)e^{i\phi (\vec{r},t)}\,,  \label{FieldEQN} \\
& \frac{\partial }{\partial t}s(\vec{r},t)=(-\gamma +i\delta _{0})s(\vec{r}%
,t)+ga(\vec{r},t)e^{-i\phi (\vec{r},t)},\,  \label{SpinEQN}
\end{align}
where we neglect the Langevin noise atomic operators by assuming that all atoms were initially and mainly remain at any later time in the ground state $\lvert 1\rangle $ and neglect the optical transition broadening in comparison with the large one photon detuning.
The signal-matter coupling in Eqs. (\ref{FieldEQN})-(\ref{SpinEQN}) is determined by the spatial chirp of the control field via the phase shift
\begin{equation}
\phi (\vec{r},t)=\left[ \vec{k}_{c}(\vec{r}_{\perp })-\vec{k}_{c0}\right]
\cdot \vec{r}-\left[ \omega _{c}(\vec{r}_{\perp })-\omega _{c0}\right]
t-\varphi _{0c}(\vec{r}_{\perp }).  \label{phase shift}
\end{equation}
In these equations $\Delta _{\perp }=\frac{\partial ^{2}}{\partial x^{2}}+\frac{\partial ^{2}}{\partial y^{2}}$, $\delta _{0}=\omega _{s}-\omega_{c0}-\omega _{3}$ is the average two-photon detuning, $\Delta =\omega_{s}-\omega _{2}$ is the one-photon detuning, $g=\frac{d_{21}\Omega }{n\Delta }\sqrt{\frac{\omega _{s}}{2\hbar \varepsilon _{0}c}}\ $ is the coupling constant responsible for the interaction between the atomic ensemble and the single-photon wave packet at the optical transition $\lvert 1\rangle $-$\lvert 2\rangle $, $\Omega =d_{32}E_{0}^{\ast }/\hbar $ is the Rabi frequency of the classical control field, $d_{ij}$ is the dipole moment of the transition between $\left\vert i\right\rangle $ and $\left\vert j\right\rangle $ states, $\gamma $ is the rate of dephasing of the Raman (spin) coherence, the decoherence rate between the levels $\lvert 2\rangle $ and $\lvert 1\rangle $ is neglected comparing with the one-photon detuning. For the sake of simplicity, the Raman transition frequency shift $|\Omega|^{2}/\Delta $ is included in the average control field frequency $\omega_{c0}$.

The performance of a quantum memory scheme can be characterized by the total efficiency $\eta $ and fidelity $\mathscr{F}^{\prime}$. The former is defined as
\begin{equation}
\eta =\frac{N_{\text{out}}}{N_{\text{in}}}\,,
\end{equation}%
where $N_{\text{in}}=\int_{-\infty }^{0}dt\,\langle a_{\text{in}}^{\dag
}(t)a_{\text{in}}(t)\rangle $ and $N_{\text{out}}=\int_{0}^{\infty }dt\times
\langle a_{\text{out}}^{\dag }(t)a_{\text{out}}(t)\rangle $. Here $a_{\text{
in}}(t)=a(z=-L/2,t<0)$ and $a_{\text{out}}(t)=a(z=\pm L/2,t>0)$, with ``$+$"
for forward retrieval and ``$-$" for backward retrieval. The fidelity $
\mathscr{F}$ can be defined as $\mathscr{F}=\eta \mathscr{F}^\prime$, where $\mathscr{F}^\prime$ is the fidelity measuring the signal preservation regardless of the efficiency:
\begin{equation}
\mathscr{F}^{\prime }=\frac{1}{N_{\text{in}}N_{\text{out}}}\left\vert
\int_{0}^{\infty }dt\,\langle a_{\text{in}}^{\dagger }(\bar{t}-t)a_{\text{out%
}}(t)\rangle \right\vert ^{2}\,,  \label{Fidelity Definition}
\end{equation}
and the retarded argument $\bar{t}-t$ takes into account the time reversal and possible time shift of the output pulse.

Equations (\ref{FieldEQN}) and (\ref{SpinEQN}) describe the storage process through the absorption of the signal field and creation of the Raman coherence (the spin wave) in the atomic ensemble. During forward retrieval the spatial chirp is reversed, so the process is described by the same equations with an opposite frequency gradient. Note that depending on the propagation direction of the control field, equations take different forms. In any case, the phase factor $e^{i\phi (\vec{r},t)}$ is crucial in the proposed quantum memory scheme for converting the single-photon wave packet into the spin waves and vice versa. For convenience, in this work we assume two-photon resonance ($\delta _{0}=0$) and consider only a linear frequency gradient across the control beam, described by the vector $\beta $, lying in the $x$-$z$ plane.

\section{Transverse excitation regime}

\label{sec: Transverse excitation}

In this section we consider the transverse excitation regime of the proposed scheme, in which the signal field propagates along the $\hat{z}$ direction and the control field propagates perpendicularly along the $\hat{x}$ direction. As a result, the frequency gradient of the control field lies along the longitudinal direction $\hat{z}$ and provides the spatial chirp $\omega _{c}(z)=\omega _{c0}+\beta z$, $z\in \lbrack -L/2,L/2]$, which is centered around the center plane $z=0$ \citep{Zhang13}. Choosing for convenience $x_{0}=t_{0}=0$, the phase shift in Eq. (\ref{phase shift}) becomes the following:
\begin{equation}
\phi (\vec{r},t)=-\beta z(t-x/c)\,,
\end{equation}
where $x\in \lbrack -R,R]$ and $R<L$.

\begin{figure}[h]
\centering
\resizebox{!}{10.7cm}{\includegraphics{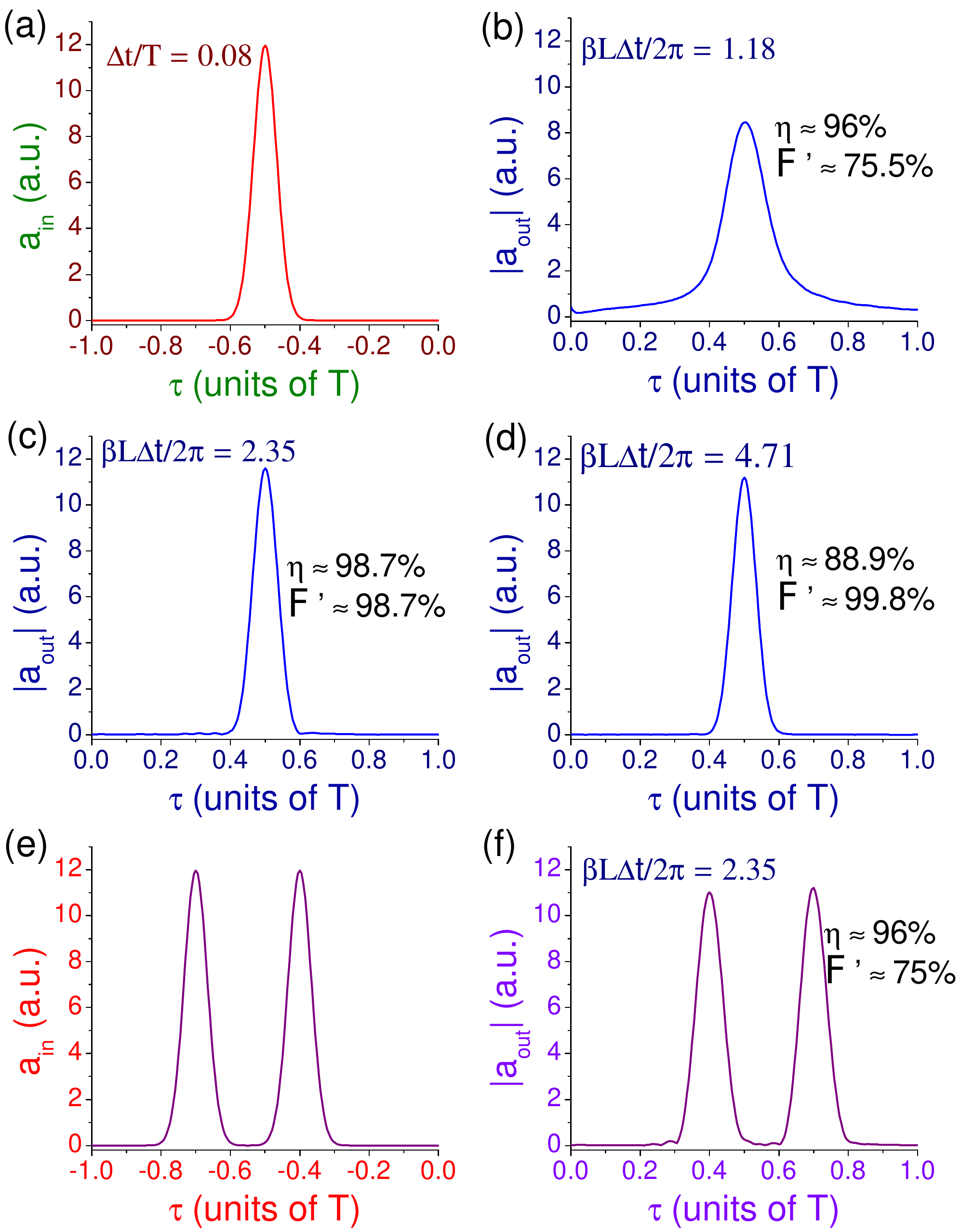}}
\caption{(Color online) (a-d): Forward retrieval signal amplitude $|a_{\text{out}}(
\protect\tau )|$ (phase modulation is not shown in the figure but taken into
account in the fidelity $\mathscr{F}^{\prime }$) as a function of the
retarded time in the transverse excitation regime for different values of
the spatial chirp $\protect\beta L$ across the control beam ((b) $\protect
2\pi \sqrt{2\ln 2} /\Delta t$, (c) $4\pi \sqrt{2\ln 2} /\Delta t$, (d) $8\pi \sqrt{2\ln 2} /\Delta t$)
for the depicted in (a) Gaussian temporal shape input signal field $a_{\text{in}}(\protect\tau )$
with a FWHM duration $\Delta t= (\sqrt{2\ln 2}/15)T$, where $T$ is the duration of the writing
process. (e-f): A double-peak Gaussian temporal shape input signal field (e) and its forward retrieval signal amplitude (f) for $\beta L = 4\pi \sqrt{2\ln 2} /\Delta t$. The graphs are plotted for the parameter $|g|^{2}NL\Delta t=13.78$. }
\label{Figure: 4PP6in1}
\end{figure}

Let us assume a short propagation time of the pulse through the medium compared to the inverse spectral broadening via spatial chirp, which, in its turn, is shorter than the pulse duration ($L/c < 2\pi/(\beta L) \lesssim \Delta t$). This allows us to neglect the retardation of the signal field, as well as the transverse dependent phase factor $\beta zx/c$. In the case of a large Fresnel number for the interaction volume from the viewpoint of the signal field, we can neglect diffraction effects and omit the transverse derivative term in Eq. (\ref{FieldEQN}). Then the equations describing the transverse excitation regime take the following form
\begin{align}
&\frac{\partial }{\partial z}a(\vec{r},\tau)=-g^{\ast }Ns(\vec{r},\tau)e^{-i\beta z\tau}\,, \label{GEMa1} \\
&\frac{\partial }{\partial \tau}s(\vec{r},\tau) =-\gamma s(\vec{r},\tau)+ga(\vec{r},\tau)e^{i\beta z\tau}\,,  \label{GEMs1}
\end{align}
where $z\in \lbrack -L/2,L/2]$ and $\tau = t-z/c$ is the local time in the signal co-moving frame. These are the same equations for the signal field operator and the spin wave in the medium as we found for the transverse scheme of the quantum storage via the control field angular scanning in Ref.~\citep{Zhang13}. Note that renaming the spin-wave amplitude as $S(\vec{r},\tau )=s( \vec{r},\tau )e^{-i\beta z\tau }$ makes Eqs. (\ref{GEMa1}) and (\ref{GEMs1}) similar to the equations describing the GEM scheme, where a controlled inhomogeneous broadening is created through the Stark or Zeeman effect by applying an external dc-electric/magnetic field with a gradient along the longitudinal direction. This analogy reveals the essence of the proposed scheme. Namely, the spatial chirp of the control field plays the same role as the longitudinal controlled inhomogeneous broadening and the temporal spectrum of the signal field is directly mapped into the longitudinal distribution of the Raman coherence. Therefore, we prove that the performance of the proposed in the present paper transverse excitation regime of the spatial-chirp scheme is the same as the performance of the GEM scheme. As an analogy to the PMC scheme, we regard the phase factor $e^{-i\beta z\tau }$ as a time dependent wave vector $\beta \tau $, which changes with time and maps the corresponding part of the signal field into a selected phase-matched spin wave. Since the frequency gradient of the control field is along $\hat{z}$, the evolution of the signal field in the course of propagation through the effectively inhomogeneous medium is complicated. The approximate analytical solution was derived in~\citep{Longdell08, Moiseev08}, and used also in~\citep{Zhang13, Clark12}. The only difference is the re-definition of the spin wave discussed above. From that solution it follows that the effective optical depth for the GEM scheme is $2\pi |g|^{2}N/\beta $, which should be larger than $1$ in order to achieve large enough efficiency. On the other hand, the longitudinal frequency chirp width $\beta L$ should be larger than the bandwidth of the signal spectrum $8\ln 2 /\Delta t$ to achieve the high fidelity, where $\Delta t$ is the full width at half maximum (FWHM) duration of the input signal. So the necessary condition for large enough efficiency and fidelity in the transverse excitation regime of the proposed spatial-chirp scheme is:
\begin{equation}
1\lesssim \frac{\beta L \Delta t}{2\pi} < |g|^{2}NL\Delta t\,.
\label{GoodConLongi}
\end{equation}
The quantum memory performance of such system was discussed in \citep{Zhang13}. One of the advantages of the GEM scheme is that the high efficiency can be achieved without backward retrieval. While this greatly simplifies the experiment, one should note that the forward retrieval does not preserve the time reversal symmetry. Although high efficiency is possible, there is a strong phase modulation of the retrieved signal, especially at the large optical depths. This is also true for the forward retrieval in the transverse excitation regime of the proposed scheme, which is done by switching the frequency gradient to the opposite $\beta \rightarrow -\beta $ and keeping the same control beam propagation direction. Numerical simulations of Eqs. (\ref{GEMa1}) and (\ref{GEMs1}) are presented in Fig. \ref{Figure: 4PP6in1}, using predictor-corrector method on spatial variable. In order to consider the ultimate performance of the proposed scheme, the decay of the spin wave during the storage time was neglected, which implies very short storage time compared to the spin wave life time. Fig. \ref{Figure: 4PP6in1} (a, c) show that very high efficiency and fidelity can be achieved for the proposed spatial-chirp scheme for the forward retrieval in the transverse excitation regime. The decrease of the fidelity in Fig. \ref{Figure: 4PP6in1} (e, f) is because the retrieved field experiences more phase modulation with reducing the distance between the pulse and the moment of the control field switching, in the same way as in the case of usual GEM~\cite{Moiseev08}. However, it is worth to note that almost perfect amplitude correlation is preserved in the retrieved field. The backward retrieval signal field, on the other hand, is free of phase modulation. It can be achieved by keeping the same frequency gradient $\beta \rightarrow \beta $ but with opposite control beam propagation direction. Because of the phase matching condition, a conjugate spin coherence should be prepared before backward retrieval by, for example, two successive non-collinearly-propagating $\pi$ pulses as discussed in~\citep{Zhang13}.

Finally, it is worth to note that the proposed all-optical GEM scheme provides some additional possibilities for manipulating the pulses, which are not available in the case of usual GEM implementation. For example, a transverse phase modulation of the read-out control field such as $\exp (i\beta z t^\prime)$ leads to the temporal shift of the output pulse on $t^\prime$ without its temporal squeezing or stretching. In other words, we can shift the pulse in time without changing the frequency gradient.

\section{Non-transverse excitation regime}

\label{sec: Non-Transverse excitation}

In a more general case, the propagating directions of the control and signal fields can make an arbitrary angle $\theta $. We make the spatial chirp centered at control beam center, then $\omega _{c}(x,z)=\omega _{c0}+\beta z\sin \theta -\beta x\cos \theta $, with $x\in \lbrack -R,R]$, $z\in \lbrack -L/2,L/2]$. Without losing generality, we take $z_{0}=x_{0}=t_{0}=0$. The phase shift in Eq. (\ref{phase shift}) is the following function of the coordinates and time:
\begin{equation}
\phi (\vec{r},t)=-\beta (z\sin \theta -x\cos \theta )\left( t-\frac{z\cos
\theta +x\sin \theta }{c}\right) .  \label{phase shift2}
\end{equation}
Consider $k_c \approx k_s$, $\Delta t \gg L/c$, $\Delta k_{sx}\sim \Delta k_{cx}\sim \beta \Delta t$, $\beta R <\beta L \sim 2\pi/\Delta t$, the transverse derivative in Eq. (\ref{FieldEQN}) can be neglected for large signal field Fresnel number $F \gg 1$. In the following we assume this condition is satisfied so that there is no transverse dynamics in the model.

Under the same assumption made in Sec. \ref{sec: Transverse excitation}, $L/c < 2\pi/ (\beta L) < \Delta t$, the $(z\cos \theta +x\sin \theta )/c$ term in Eq. (\ref{phase shift2}) can be neglected and we are able to approximate the phase shift in Eq. (\ref{phase shift}) by $\phi (\vec{r},t)=-\beta(z\sin \theta -x\cos \theta )\,t$. Defining a variable $S(\vec{r},t)=s\left(\vec{r},t\right) e^{-i\beta (z\sin \theta -x\cos \theta )t}$, then we get the quantum memory equation for non-transverse excitation regime as
\begin{align}
\frac{\partial }{\partial z}a\left( z,\tau ;x\right) & =-g^{\ast }NS\left(
z,\tau ;x\right) \,,  \label{GEMa2} \\
\frac{\partial }{\partial \tau }S\left( z,\tau ;x\right) & =-[\gamma +i\beta
(z\sin \theta -x\cos \theta )]S\left( z,\tau ;x\right)  \notag \\
& +ga\left( z,\tau ;x\right) \,,  \label{GEMs2}
\end{align}
where the retarded effect is neglected. These are the same equations as in the angular scanning scheme, analyzed in~\cite{Zhang13}. Therefore these two schemes provide the same performance. The forward and backward retrieval can be realized in the same manner as it was discussed above in the case of transverse propagation. Other possibilities, such as retrieving at the control field angle equal to $-\theta $ or $\pi -\theta $ ($\theta \neq 0$), require external spin wave vector rotation  before retrieval. Decreasing the angle $\theta $, below the critical one: $\beta L \sin(\theta) \Delta t/(2\pi) <1 $ reduces the longitudinal absorption window, making it too small to accommodate the full Fourier components of the signal.

\section{Experimental issues}

\label{sec: Spatial chirp of Gaussian control beams}
Quantum memory based on the proposed approach requires en ensemble of atoms stationarily distributed in space. The Raman configuration calls for a rather strong optical transition. Based on these two requirements, nitrogen vacancy (NV) centers in diamond is a promising material for realization of the suggested scheme. NV diamond has much stronger electro-dipole Raman transitions compared to rare-earth ions. By mixing ground or excited states via magnetic field or electric field/transverse strain, different $\Lambda$-level structures have been employed in the experimental demonstrations of EIT in NV ensemble~\cite{Hemmer01, Acosta13}. Under low strain, another $\Lambda$-level structure is used for spin-photon entanglement experiment~\cite{Togan10}. The recent theoretical proposal~\cite{Heshami13} is of particular relevance for us since it proves a feasibility of the Raman quantum memory protocol based on modulation of the control field. In this case, high external static electric field and low magnetic field are applied to create polarization selection rules. The spin triplet ground levels $|S=1,m_s = \pm 1\rangle$ are mixed into two new ground states $|\pm\rangle$, together with spin triplet excited state split by transverse electric field ~\cite{Batalov09}, forming a $\Lambda$-level system~\cite{Heshami13}. Another candidate is silicon vacancy (SiV) in diamond. It has similar oscillator strength as NV center~\cite{Hepp14} but larger Debye-Waller factor at $738$~nm~\cite{Neu11}. However, its spin coherence lifetime is yet to be studied.

Optical fields with a spatial chirp across the beam have been widely used in Fourier synthesis pulse shaping ~\citep{Heritage85}, multiphoton microscopy \citep{Zhu05,Oron05}, micromachining~\citep{Vitek10}, etc., and have also been proposed to assist the achievement of large EIT bandwidth by applying a magnetic field transverse gradient~\citep{Sun05}. They commonly exist in the inhomogeneous dispersive medium, and can be produced by means of prism or grating diffraction~\citep{Durfee12}, pulse spatial modulation~\cite{Futia11}, etc.. Depending on the spectrum width of the chirp and control beam duration, the proposed scheme operates at different regimes.

In the regime of sub-microsecond-pulse quantum storage, the spatial chirped beam can be prepared by, for example, frequency modulated reticles~\cite{Sanders91}. In this technique the bandwidth of the control beam will be determined by the pattern, size and the spinning speed of the reticle. For intensity modulation, one of the two side bands should be blocked for the current application. Assuming the pattern has a reciprocal wave vector $\Delta k=1000$~mm$^{-1}$,  input beam size (or the radius of the reticle, whichever is smaller) $1$~cm, and spinning speed $500$~Hz~\cite{Futia11}, we have $\beta=3.14\times 10^9$~s$^{-1}$m$^{-1}$. According to condition (\ref{GoodConLongi}), a medium with $|g|^2N = 3\times 10^{9}$~s$^{-1}$m$^{-1}$ can store and retrieve $\sim 200$~ns signal pulse with high efficiency and fidelity. This can be fulfilled if NV diamond is used as storage medium. The inhomogeneous broadening of the optical transition in the whole ensemble of NV centers can be on the order of $10$~GHz~\cite{Santori06} at the NV density $8\times 10^{15}$ cm$^{-3}$~\cite{Heshami13}. The dipole moment corresponding to zero phonon line can be estimated as $\sim 2.3\times 10^{-30}$~Cm based on oscillator strength $\sim 0.1$ and Debye-Waller factor $0.035$. We take the $\Lambda$-level configuration recently suggested in~\cite{Heshami13} for realization of the traditional protocol of Raman quantum memory based on modulation of control field amplitude. Following ~\cite{Heshami13}, let us consider a sub-ensemble of NV with an effective inhomogeneous broadening $0.2$~GHz with density $N=2\times 10^{13}$~cm$^{-3}$. Such sub-ensemble is to be produced by the hole burning and repumping technique. The spectral hole is characterized by sharp edges and small background absorption. Particularly, in considered scheme~\cite{Heshami13} a spectral hole of the width $2.87$~GHz can be created by making $|+\rangle$ and $|-\rangle$ states empty (all NV centers are transferred into the $|0\rangle$ state). Then the $|+\rangle$ state is re-populated by optical pumping, which produces an inhomogeneously broadened optical transition with a narrow linewidth (limited from below by the homogeneous width of the optical transition) inside the spectral hole. The effective reduction of the inhomogeneous linewidth leads to proportional reduction of the effective density to the required value (so that the spectral density remains about the same). The produced narrow absorption line does not need to be Gaussian and may have the sharp edges, which allows us to use rather small one-photon detuning, $0.5$~GHz. In its turn, as it is shown below, using of such small detuning helps to suppress noise originating in the considered $\Lambda$-level scheme~\cite{Heshami13} from the off-resonant scattering of the control field via closely lying (separated by $1.5$~GHz) excited state (due to the specific selection rules). The required $|g|^2N$ is fulfilled with control field intensity $\sim 140$~W/cm$^{2}$. The intensity up to $280$~W/cm$^2$ was used for demonstrating EIT in NV ensemble~\cite{Hemmer01}, and much higher intensities are used in single NV experiments. It should be noted that the noise can also be reduced by frequency filtering. In doing so, larger values of $\Delta$ are possible. A cavity-assisted Raman interaction may be helpful for reducing the control field power. In such a case, it becomes possible to implement the proposed scheme in some rare-earth doped crystals like Er$^{3+}$ or Nd$^{3+}$ in YLiF$_4$~\citep{Zhang14LP}.

The rate of spontaneous Raman scattering of the control field into a specific spatial mode (corresponding to the signal field) can be estimated as $|g^\prime|^2 NL$~\cite{Raymer81}, which is equivalent to $|g|^2NL(\Delta/\Delta^\prime)^2|d_{21}^\prime /d_{21}|^2|d_{32}^\prime /d_{32}|^2$, where $\Delta$, $d_{21}$, $d_{32}$ are the control field one-photon detuning and dipole moments in the working $\Lambda$ scheme, and $g^\prime$, $\Delta^\prime$, $d_{21}^\prime$, $d_{32}^\prime$ are the corresponding quantities in the noise channel. The total number of scattered photons during storage and retrieval processes each of duration $T$ is consequently equal to $2|g|^2NLT(\Delta/\Delta^\prime)^2|d_{21}^\prime /d_{21}|^2|d_{32}^\prime /d_{32}|^2$, which should be made smaller than $1$ to neglect spontaneous Raman scattering noise. In the considered case of the NV diamond Raman configuration~\cite{Heshami13}, substituting the values of the corresponding parameters ($|g|^2NL = 3\times 10^{7}$~s$^{-1}$, $\Delta^\prime /2\pi = \Delta/2\pi+1.5$~GHz, taking for simplicity $d_{32}^\prime = d_{32}$ and $d_{21}^\prime = 0.75d_{21}$), we find the last condition fulfilled by $\Delta/2\pi<0.6$~GHz, proving that signal-to-noise ratio remains greater than $1$ for storage and retrieval of a single photon in a total time window $>700$~ns.

In the sub-nanosecond-pulse regime, let us consider the implementation of the spatial chirp via a spatial dispersion of the Gaussian control beam~\citep{Gu04}. Taking the transverse excitation regime as example, and assuming the higher order chirp is absent~\citep{Durst08}, then the control beam with Gaussian spectrum and Gaussian transverse profile can be written in the frequency domain as
\begin{equation}
E_{c}(z,\omega )=\mathscr{E}_{0}e^{-\left( \frac{\omega -\omega _{c0}}{
\Delta \omega _{c}}\right) ^{2}}e^{-\left( \frac{z-\zeta (\omega -\omega
_{c0})}{\Delta \mathrm{w}}\right) ^{2}},  \label{EcExpSpectrum}
\end{equation}
in which $2\sqrt{\ln 2}\Delta \omega _{c}$ is the FWHM spectrum bandwidth, $2\sqrt{\ln 2}\Delta \mathrm{w}$ is the FWHM spatial width of each spectrum component, $\zeta $ is the spatial dispersion, and the propagation (along $\hat{x}$) effect is neglected. The inverse Fourier transform of (\ref{EcExpSpectrum}) gives the control beam in time domain:
\begin{equation}
E_{c}(z,t)=\frac{\Delta \omega _{c}\mathscr{E}_{0}}{\sqrt{2}\kappa }e^{-
\frac{\Delta \omega _{c}^{2}t^{2}}{4\kappa ^{2}}}e^{-\frac{z^{2}}{\kappa
^{2}\Delta \mathrm{w}^{2}}}e^{-i\frac{\Delta \omega _{c}^{2}\zeta }{\kappa
^{2}\Delta \mathrm{w}^{2}}zt}e^{-i\omega _{c0}t}, \label{Eczt}
\end{equation}
where the control beam is both transversely expanded and temporally stretched by a factor $\kappa =\sqrt{\Delta \mathrm{w}^{2}+\zeta ^{2}\Delta \omega _{c}^{2}}/\Delta \mathrm{w}$ comparing with a beam without spatial dispersion ($\zeta =0$), and the frequency gradient is $\beta =\Delta \omega_{c}^{2}\zeta /(\kappa ^{2}\Delta \mathrm{w}^{2})$. The term $e^{-i\beta zt}$ indicates that the control beam phase front rotates while the field is propagating.

For the quantum memory application, at least three conditions should be fulfilled: (i) spatial coverage: $2\sqrt{\ln 2}\kappa \Delta\mathrm{w}>L$; (ii) temporal coverage: $4\sqrt{\ln 2}\kappa /\Delta \omega_{c}>\Delta t$; and (iii) spectral coverage: $\beta L=\Delta \omega_{c}^{2}\zeta L/(\kappa ^{2}\Delta \mathrm{w}^{2})>2\pi /\Delta t$, where $L$ is the medium length, $\Delta t$ is the signal field duration. Besides, the $z$-dependence of the control field amplitude should be sufficiently smooth so that the spatial chirp produced by the Raman frequency shift $\frac{|\Omega _{0}|^{2}}{\Delta }(1-e^{-\frac{2z^{2}}{\kappa ^{2}\Delta \mathrm{w}^{2}}})$, where $\Omega _{0}=\Omega(z=0)$, would be negligible compared with the spatial chirp $|\beta z|$. Indeed, since $\kappa\Delta \mathrm{w}\gg L\geqslant z$, taking into account condition (\ref{GoodConLongi}) we get the ratio of the Raman shift and $|\beta z|$ approximately equal to $\frac{|\Omega _{0}|^{2}}{\beta \Delta }\frac{2|z|}{\kappa ^{2}\Delta \mathrm{w}^{2}} \ll \frac{2|z|}{\kappa \Delta \mathrm{w}} \frac{|\Omega_0^2|}{\Delta} / \frac{2\pi}{\Delta t} \ll 1$.

As one of the simplest schemes to generate control beam of the form Eq. (\ref{Eczt}) fulfilling the conditions (i) - (iii), let us consider specifically a grating-lens pair with the grating and the medium placed at the back and front focal planes of the lens~\cite{Wefers95}. The switching of the spatial dispersion can be achieved by adding two additional lenses in a 6-F configuration. Take $\lambda_{c0}=700$~nm and the focal length $f=8$~m and assume the angular dispersion of the grating is $d\theta /d\lambda =2.5\times 10^{-3}$~nm$^{-1}$, the spatial dispersion $\zeta = -\frac{\lambda_{c0}^2}{2\pi c} f \frac{d\theta}{d\lambda}\big|_{\lambda_{c0}} = -5.2\times 10^{-15}$~sm. For a $29$~ps seed Gaussian pulse and $\Delta \mathrm{w}=10$~$\mu$m focal spot size of each individual frequency component, we get the frequency gradient $\beta=-1.9\times 10^{14}$~m$^{-1}$rad/s, and the stretching factor $\kappa=100$ which expands the transverse size of the control beam to $1$~mm and stretches the duration to $2.9$~ns. This corresponds to a sub-nanosecond-pulse quantum storage with signal field duration in between of $29$~ps and $2.9$~ns. The medium length $L<1$~mm makes sure that the scheme operates at the long pulse regime. The difficulty of sub-nanosecond-pulse quantum storage is that a high coupling constant and atomic number density are required. In the present example, the medium has to demonstrate $|g|^2N > |\beta/2\pi| = 3\times 10^{13}$~s$^{-1}$m$^{-1}$. Such coupling constant and atomic number density are not yet feasible in typical NV centers. With the development of material science we hope color centers (such as SiV) in diamond and other materials with much higher oscillator strength and atomic density and much smaller optical inhomogeneous broadening could become available for such sub-nanosecond-pulse quantum storage based on Raman interaction.

In principle, the gap between the above two regimes can be filled by generating chirped beam with $\sim$GHz spectrum width through, e.g., acousto-optic modulator (AOM). In such a case, a number of sub-beams are split from the original one, each with different central frequency due to AOM modulation, and aligned side-by-side to form a spatial-chirp control beam. A significant difference is that, if the number of the sub-beams is small, the frequency gradient becomes discrete, thereby forming an optical frequency comb. Contrary to AFC, such an optical comb is not limited by the inhomogeneous broadening of the atomic levels, can be easily created without background absorption and modified between storage and retrieval. These features make the proposed scheme promising from the viewpoint of experimental realization. The performance of quantum memory subject to discrete frequency gradient will be discussed elsewhere.

The storage time is limited by homogeneous and inhomogeneous broadening of the spin transition, which in the case of NV diamond can be megahertz or smaller \cite{Acosta13,Wilson05}. However, the inhomogeneous broadening of the spin transition can be effectively eliminated using the microwave pulse sequences (see, e.g., \cite{Heinze13}). As a result, the storage time is determined by the homogeneous linewidth of the two-photon transition, which may be of the order of kilohertz under cryogenic temperature. The spin coherence time up to $0.6$~s is demonstrated using decoupling techniques at $77$~K~\cite{Bar-Gill13}.

\section{Conclusions}

\label{sec:Discussion and Conclusion}

In the photon echo based quantum memory schemes, the essential part is to produce a re-phase of the dipole moments. The classical photon echo, for example, the two-pulse photon echo, uses the strong $\pi $ pulses to produce the re-phase, thus introducing a noise due to spontaneous emission and amplification originated from a population inversion, which makes it not suitable for the quantum field storage~\citep{Ruggiero09}, unless techniques such as silent echo~\citep{Damon11} is used. Instead, the CRIB, and in particular the GEM schemes create the re-phase by manipulating an effective inhomogeneous broadening and introduces no population inversion. In order to achieve this, dc-Stark or Zeeman effect should be used. In order to avoid complications due to required manipulations with the medium for the creation of the inhomogeneous spectral broadening structures, for example, by applying the external electric or magnetic fields with the spatial gradient, and for the material systems which do not demonstrate the dc-Stark or Zeeman effects, the alternative ways of creating effective controllable inhomogeneous broadening should be adopted.

In the present scheme, we create an effective controllable inhomogeneous broadening of the Raman transition by a spatially inhomogeneous control field pattern with a frequency gradient across the control beam, that is the so-called spatial chirp. Since the control field wave pattern mimics the longitudinal inhomogeneous broadening, the atomic medium acts like a recorder which
stores the spectrum components of the input signal as in GEM scheme.
Meanwhile in terms of the phase matching control, a time dependent transverse wave vector $\beta \tau $ created by the control field wave front rotation continuously records the signal field by the corresponding phase-matched spin wave components. Eventually the temporal shape of the signal field maps into the spatial distribution of the spin wave. In forward retrieval, reversing the frequency gradient of the control field has similar effect as reversing the dc-field gradient in the GEM scheme or reversing the phase matching sweeping in the PMC scheme.

By combining the basic ideas of the GEM scheme and the off-resonant Raman interaction based scheme in the proposed spatial-chirp scheme, we naturally come to the new memory scheme which contains the advantages of both, while, unlike the three-level GEM scheme, any direct manipulation of the resonant medium is avoided. As far as a practical implementation is concerned, any material demonstrating strong Raman transitions and long Raman coherence time can be employed. In particular, nitrogen-vacancy centers and silicon vacancy centers in diamond seem to be the most promising systems.

In summary, we propose an all-optical quantum memory scheme based on the off-resonant Raman interaction of the signal quantum field with a homogeneous atomic medium illuminated by a constant-in-time control field with a spatial chirp of frequency across its beam. Its mathematical model is similar to the GEM scheme. However, physically the proposed scheme is completely different and has very important experimental and implementation advantages compared to the controllable-reversible-inhomogeneous-broadening based and other schemes. These advantages originate from its all-optical nature. Unlike the Raman or phase-matching-control (PMC) schemes, the proposed spatial-chirp one does not require, in principle, any synchronization or temporal manipulation of the control field and the atomic medium. Unlike the GEM scheme, it does not require the atomic medium possessing the Stark or Zeeman effect and the use of the external electric or magnetic fields.

\begin{acknowledgments}
The authors thank A. Sokolov for very useful discussions. We gratefully acknowledge support of the National Science Foundation (Grant No. PHY-$1307346$), RFBR (Grant No. 12-02-00651-a), and the Program of the Presidium of RAS ``Quantum mesoscopic and disordered structures". X.Z. is partially supported by the Herman F. Heep and Minnie Belle Heep Texas A$\&$M University Endowed Fund held/administered by the Texas A$\&$M Foundation.
\end{acknowledgments}

{} 
\bibliographystyle{apsrev4-1}
\bibliography{QMSC}

\end{document}